\documentstyle[11pt]{article}
\begin{document}
\newcommand{\eqn}[1]{(\ref{eq:#1})}
\newcommand{\eq}{\begin{equation}}
\newcommand{\en}{\end{equation}}
\newcommand{\bea}{\begin{eqnarray}}
\newcommand{\eea}{\end{eqnarray}}
\newcommand{\nn}{\nonumber \\ }
\newcommand{\bdm}{\begin{displaymath}}
\newcommand{\edm}{\end{displaymath}}
\newcommand{\A}{\cal{A}_\gamma }
\newcommand{\At}{\tilde{\cal{A}}_{\gamma}}
\newcommand{\Ak}{{\cal A}_{3k}}
\newcommand{\SU}{\widehat{SU(2)}}
\newcommand{\SP}{\widehat{SU(2)^+}}
\newcommand{\SM}{\widehat{SU(2)^-}}
\newcommand{\SSS}{\widehat{SU(3)}_{\tilde{k}^+}}
\newcommand{\br}{\langle}
\newcommand{\kt}{\rangle}
\newcommand{\bra}[1]{\langle {#1}|}
\newcommand{\ket}[1]{|{#1}\rangle}
\newcommand{\lm}{\ell^-}
\newcommand{\lp}{\ell^+}
\newcommand{\la}{\lambda}
\newcommand{\al}{\alpha}
\newcommand{\eps}{\epsilon}
\newcommand{\vl}{\vec{\lambda}}
\newcommand{\pa}{\partial}
\newcommand{\ktp}{\tilde{k}^+}
\newcommand{\ktm}{\tilde{k}^-}

\newcommand{\NP}[1]{Nucl.\ Phys.\ {\bf #1}}
\newcommand{\Prp}[1]{Phys.\ Rep..\ {\bf #1}}
\newcommand{\PL}[1]{Phys.\ Lett.\ {\bf #1}}
\newcommand{\NC}[1]{Nuovo Cim.\ {\bf #1}}
\newcommand{\CMP}[1]{Comm.\ Math.\ Phys.\ {\bf #1}}
\newcommand{\PR}[1]{Phys.\ Rev.\ {\bf #1}}
\newcommand{\PRL}[1]{Phys.\ Rev.\ Lett.\ {\bf #1}}
\newcommand{\MPL}[1]{Mod.\ Phys.\ Lett.\ {\bf #1}}
\newcommand{\IJMP}[1]{Int.\ J.\ Mod.\ Phys.\ {\bf #1}}
\newcommand{\JETP}[1]{Sov.\ Phys.\ JETP {\bf #1}}
\newcommand{\TMP}[1]{Teor.\ Mat.\ Fiz.\ {\bf #1}}
\def\spinst#1#2{{#1\brack#2}}
\begin{titlepage}
\null
\begin{flushright}
DFTT-61/92 \\
October, 1992
\end{flushright}
\vspace{1cm}
\begin{center}
{\Large\bf
Some Universal Features of the Effective String Picture \\
of Pure Gauge Theories \footnote{Talk given by F.Gliozzi at LATTICE'92}
\par}
\vskip 3em
\lineskip .75em
\normalsize
\begin{tabular}[t]{c}
{\bf M.Caselle, F. Gliozzi and S.Vinti}\\ \\
{\sl Dipartimento di Fisica Teorica dell'Universit\`a di Torino} \\
{\sl and INFN, Sezione di Torino}\\
{\sl Via P. Giuria 1}, {\sl I-10125 Torino, Italy}
\end{tabular}
\vskip 1.5em
{\bf Abstract}
\end{center} \par
The effective string describing the large distance behaviour of the
quark sources of gauge field theories in the confining phase in D=3 or
D=4 space-time dimensions can be formulated as a suitable 2D conformal
field theory on surfaces with quark loops as boundaries. Some universal
relations among the string tension , the thickness of the colour
flux tube , the deconfinement temperature and a lower bound of the
glueball mass spectrum are discussed.

\vskip.3 cm
\end{titlepage}
\baselineskip=0.8cm
%\renewcommand{\thefootnote}{\arabic{footnote}}
%\setcounter{footnote}{0}
%%%%%%%%%%%%%%%%%%%%%%%%%%%%%%%%%%%Q

\def\c{~\cite}
\def\eq{\begin{equation}}
\def\en{\end{equation}}
\def\ss{\sqrt{\sigma}}
\def\Tc{T_c/\ss}
\def\mg{m_G/\ss}
\def\tm{T_c/m_G}
\def\sr{\ss R_f}
\def\PL{Phys. Lett.}
\def\NP{ Nuc. Phys.}
\def\Z{\hbox{\rm Z{\hbox to 3 pt{\hss\rm Z}}}}
\newcommand{\ttbs}{\char'134}
\newcommand{\AmS}{{\protect\the\textfont2
  A\kern-.1667em\lower.5ex\hbox{M}\kern-.125emS}}

% add words to TeX's hyphenation exception list
%\hyphenation{financial created another}

% declarations for front matter

%\maketitle

\section{Introduction}

The colour flux joining a pair of quarks in the confining phase is
concentrated inside  a  tube of small but finite thickness. It is
generally believed\c{no} that this thin tube behaves like a
vibrating string when these quarks are pulled very far apart.

By combining these two simple properties with the L\"uscher\c{lu}
description of the flux tube beyond the roughening, we derive,
using  a simple argument on the behaviour of
 horizontal Wilson loops at high temperature, a general
relation between the deconfinement point $T_c$ and the string tension
$\sigma$ which is universal $i.e.$ it does not depend on the specific
choice of the gauge group, being only a function of the dimension $D$
of the space-time. This relation fits nicely  the existing data of
numerical simulations with pure lattice gauge theories in $D=3$
dimensions as well as the new data for $SU(2)$ in $D=4$, while the
$SU(3)$ value seems at present a bit different.

Assuming further that the effective string cannot self-overlap, as
suggested by the strong coupling expansion of 3D Ising  gauge model, we
argue, using some recent results\c{david,sed} on self-avoiding random
surfaces, that this string is described, in the infrared limit and at
zero temperature, by a 2D conformal field theory (CFT) with $D-2$ free
bosons  compactified on a circle with a specific value of the radius.
We argue that such a compactification radius approximately measures
the thickness of the colour flux tube.

The effective string picture
arising in this way coincides with that proposed some time ago
\c{noi2,noi3} in order to fit accurately the numerical data on the
expectation value of the Wilson loops for various gauge systems in
three and four space-time dimensions and based on a different
argument\c{cb}.

The property of non-overlapping of the colour flux tube can also be used
to obtain a universal lower bound on the mass spectrum of the glueball
which is numerically well verified by the mass of the $0^+$ glueball
state both in $SU(2)$ and $SU(3)$ 4D gauge systems .
%%%%%%%%%%%%%%%%%%%%%%
\section{ Transition Temperature}
Large space-like or {\sl horizontal} Wilson loops ({\sl i.e.} loops
defined on a constant time slice) in a gauge system
at a nonzero temperature cannot be used as order parameter of
deconfinement. In particular, at high temperature, they may show area
law behaviour without static quarks being confined\c{sve}. Nevertheless
 they provide us with some important information on the transition to
the the gluon quark plasma.
Consider  indeed the horizontal  Wilson loop in a 3D gauge system at a
temperature $T=1/L$ as drawn in fig.1.
Owing to the periodic boundary conditions along the imaginary time, the
field $x_\perp$ describing the displacements  of the effective string
joining the $q\,\bar q$ sources
 is obviously compactified  on a circle of length $L=1/T$, $i.e.$
\eq
x_\perp\equiv x_\perp +L~~~.
\label{tra}
\en
$x_\perp$ may be considered as the Goldstone mode related to the
spontaneous symmetry breaking of the translational invariance along the
time direction\c{lu}.

\vskip.3cm
\begin{center}
\begin{picture}(180,90)(0,0)
\put(0,40){\line(1,0){120}}
\put(0,0){\line(1,0){120}}
\put(50,90){\line(1,0){120}}
\put(50,50){\line(1,0){120}}
\put(120,0){\line(1,1){50}}
\put(120,40){\line(1,1){50}}
\put(120,0){\line(0,1){40}}
\put(170,50){\line(0,1){40}}
\put(0,0){\line(0,1){15}}
\put(0,20){$\frac{1}{T}$}
\put(0,25){\line(0,1){15}}
\put(0,40){\line(1,1){50}}
\put(0,0){\line(1,1){50}}
\put(50,50){\line(0,1){40}}

\put(25,20){\line(1,0){70}}
\put(25,20){\line(1,1){25}}
\put(85,20){\line(1,1){25}}
\put(50,45){\line(1,0){70}}

\put(35,20){\line(1,1){25}}
\put(45,20){\line(1,1){25}}
\put(55,20){\line(1,1){25}}
\put(65,20){\line(1,1){25}}
\put(75,20){\line(1,1){25}}
\put(32,27){\line(1,0){70}}
\put(38,33){\line(1,0){70}}
\put(43,38){\line(1,0){70}}
\put(95,20){\line(1,1){25}}

%\put(26,21){\line(1,0){60}}
%\put(27,21){\line(1,1){25}}
%\put(84,21){\line(1,1){25}}
%\put(51,46){\line(1,0){60}}

\end{picture}
\end{center}
%\end{figure}
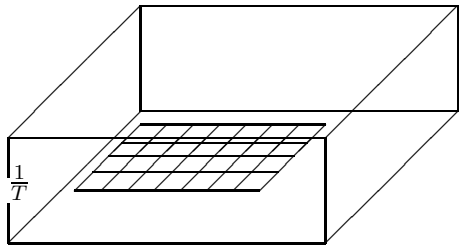
{\sl Horizontal Wilson loop at finite $T$ bordering a surface of
minimal area.}
\vskip .3cm
Denoting by $W_h(R,R')$ an horizontal rectangular Wilson loop of size
$R\times R'$, the  free energy $F_T=-\log{\langle W\rangle}$ of the
associated $q\,\bar q$ pair has the following asymptotic expansion
\eq
\lim_{R'\to\infty} \frac{F_T(R,R')}{R'}=\sigma_hR-
\frac{\tilde{c}\pi}{24R}+\dots
\label{ft}
\en
where $\sigma_h$ is the horizontal string tension (which is different
from the force experienced by a $q\,\bar q$ pair in the time direction)
and the $1/R$ term is the L\"uscher universal term, {\sl i.e.} the
zero-point energy contribution  as it arises in the
CFT of a strip (the world-sheet of the effective string) with the
quark pair as boundaries: $\tilde c=c-24h(T)$ is the effective central
charge and $h(T)$ is the conformal weight of the minimal string state
propagating along the strip. In the CFT obeyed by $x_\perp$ one has $c=1
$ ($c=D-2$ for D space-time dimensions ) and the spectrum of the allowed
conformal weights  is a known function of $T=1/L$.

Increasing the temperature of the system reduces the phase space of the
colour flux tube until it fills the whole space. At this point the flux
tube begins to be squeezed between the two opposite sides of the temporal
box; there will be a critical value  $T_c$, later identified with
the deconfinement temperature,  such that for
$T\ge T_c$ the distribution of the colour flux  along the temporal
axis becomes uniform. As a consequence, the translational invariance in
the time direction is restored, hence the Goldstone field $x_\perp$
describing the string fluctuations disappears.

It follows that the zero-point energy of the effective string  must
vanish {\sl i.e.} $\tilde c=0$, which it is possible only if
there is in the conformal spectrum a state of weight
\eq
h(T_c)={c}/{24}~~~.
\label{hc}
\en
Note that $\tilde c$, being proportional to zero-point energy,
measures the number of local degrees
of freedom of the CFT. Its vanishing tells us that at the deconfining
point the effective string theory has at most a discrete set of degrees
of freedom, $i.e.$ it behaves like a topological conformal field theory
(TCFT).
Actually most TCFT's may be formulated as (twisted)
$N=2$ superconformal theories (SCFT). It turns out
that in any $N=2$ SCFT there is a physical state of conformal weight
$h=c/24$ . Conversely one is led to conjecture
that any CFT with a weight $h=c/24$ is promoted to a $N=2$ SCFT. This
is almost trivially true for $c=1$ and $c=2$, $i.e.$ $D=3$ and $D=4$,
which are the cases we are interested in.
In particular, in the whole set of $c=1$ CFT's that can be
written in terms of the field $x_\perp$ compactified on a circle of
arbitrary radius,
the only theory with  a state of conformal weight $h=\frac{1}{24}$ is
precisely the one where the conformal symmetry is promoted to a N=2
extended supersymmetry. This allows us to select a compactification
radius and hence a specific value for $T_c$\c{tc}:
\eq
\frac{T_c}{\sqrt\sigma}=\frac{\sqrt{3}}{\sqrt{(D-2)\pi}}~~~.
\label{tc}
\en
The $D$ dependence has been inserted to take into account also the
other  interesting case of $D=4$, which can be treated in the same way
\c{app}.

Remarkably enough, our determination of $T_c$  coincides with the
Hagedorn temperature and with the value predicted for the Nambu-Goto
string. Our argument suggests that this temperature is universal and
does not depend on the gauge group.
\section{Self-avoiding String}
The strong coupling expansions of any gauge theory show that the
colour flux tube cannot self-overlap freely as Nambu-Goto string would
imply, but must obey some constraint which depends on the gauge group.
The simplest case is the $\Z_2$ gauge group where this constraint tells
us that the flux tube is self-avoiding.

Actually it has been shown
\cite{david}  the exact equivalence between a model of random
self-avoiding surfaces embedded in a three-dimensional lattice and a
$O(N)$ lattice gauge theory for any $N$. Moreover this model
has a phase transition belonging to
the same universality class of the $\Z_2$ gauge model.

This suggests assuming that the effective string
is described at least approximately, for any gauge group, by a
self-avoiding string.

It is well known that the theory of random surfaces can be formulated in
terms of a 2D quantum gravity coupled to some matter fields describing
the embedding of the surfaces in a target space. In the case of a 3D
euclidean space, these matter fields are of course the
three coordinates $x_i(\varsigma,\tau)$,  $(i=1,2,3)$, describing the
embedding of the surface as a function of the world-sheet parameters
$\varsigma$ and $\tau$. Their contribution to the central charge is $c_x=
3$. The self-avoiding property induces further conformal matter which
controls the self-intersections of the surface. There are two kinds of
such matter fields for 3D embedding\c{sed}: the lines of self-
intersection may be associated to a free fermion $\psi$ with
$c_\psi=1/2$ while the end-points of these lines, which are topological
defects of the embedded surface, are described by an anyon $\chi$
with $c_\chi=-11/2$; then
$c_{matter}=c_x+c_\psi+c_\chi=-2$. As an aside remark, note that this
is exactly the central charge of
the conformal matter describing a self-avoiding polymer in the dense
phase; the deconfinement point, where the effective string
becomes N=2 supersymmetric, may then be interpreted as the tricritical
$\Theta$ point of the polymer\c{saw}.

For very large interquark distances there are reasons to believe that
the effective string is described only in terms of matter conformal
fields. Actually this is what happens in the transverse gauge\c{olesen}
where, as we already anticipated, the asymptotic string is described,
for D=3, by a free bosonic field $x_\perp$ with $c=1$. Consistency with
the above-mentioned theory of random surfaces implies that there should
be a $c=1$ CFT equivalent to the one with $c=-2$. Surprising as it may
be, this equivalence exists and is unique\c{df}: with suitable boundary
conditions\c{saw}  the partition function for the $c=-2$ CFT can be
expressed in
terms of a gaussian model where the free field $x_\perp$ is compactified
like in  eq.(1), but with a compactification length $L'=2R_f$
slightly smaller then that associated to the (inverse) of critical
temperature, indeed one finds\c{saw,app}
\eq
\sqrt{\sigma}R_f={\sqrt{\pi(D-2)}}/{4}~~.
\label{rf}
\en
In analogy with the string picture at the deconfinement point, we
think that $R_f$  measures  the transverse radius of the colour flux
tube at zero temperature. Notice that this gaussian model with precisely
such a compactification length has been selected among various possible
descriptions of the asymptotic effective string as the one which fits
better to the data of numerical simulations in many different gauge
systems\c{noi2,noi3,cb}.
\begin{table*}[hbt]
% space before first and after last column: 1.5pc
% space between columns: 3.0pc (twice the above)
\setlength{\tabcolsep}{0.5pc}
% -----------------------------------------------------
% adapted from TeX book, p. 241
\newlength{\digitwidth} \settowidth{\digitwidth}{\rm 0}
\catcode`?=\active \def?{\kern\digitwidth}
% -----------------------------------------------------
\caption{Comparing string values with data from LGT simulations}
\label{tab:table}
\begin{tabular}{lrrrrrr}
\hline
                 & \multicolumn{3}{l}{3D Gauge Systems}
                 & \multicolumn{3}{l}{4D Gauge Systems} \\
\cline{2-4} \cline{5-7}
                 & \multicolumn{1}{r}{string}
                 & \multicolumn{1}{r}{$\Z_2$}
                 & \multicolumn{1}{r}{$SU(2)$}
                 & \multicolumn{1}{r}{string}
                 & \multicolumn{1}{r}{$SU(2)$}
                 & \multicolumn{1}{r}{$SU(3)$} \\
\hline
$\Tc$    & $ 0.977$ & $1.17(10)$\c{tc} & $ 0.94(3)$\c{fp}
         & $ 0.691$ & $0.69(2)$\c{fh}  & $ 0.56(3)$\c{fh}\\
$\mg$    & $ 2.596$ &~&$4.77(5)$\c{te}& $3.671$ & $3.7(2)$\c{vb}
         & $3.5(2)$\c{vb}\\
$\tm$    & $ 0.376$ &~&~& $0.188$ & $0.180(16)$\c{fh}
         & $0.176(20)$\c{fh}\\
$\sr$    & $ 0.443$ &~&~& $0.627$ & $0.4\div0.6$\c{dg}& ~\\
\hline
\end{tabular}
\end{table*}

The idea that the asymptotic effective string is described by a
compactified boson allows us also to get a lower bound for the
glueball mass.
This spectrum can be evaluated by studying the exponential decay of the
correlation function of small quark loops at large distance. If $\gamma$
is a small circular loop with center on a point $x$ and $W_x(\gamma)$
is  the  associated Wilson loop operator, we have, asymptotically
\eq
\langle W_x(\gamma)W_{x+L}(\gamma)\rangle\;\sim\;{\rm e}^{-m_GL}~~~,
\label{glueball}
\en
where $m_G$ is the mass of the lowest glueball. In a string picture this
expectation value can be written as the partition function of a CFT on a
surface with the topology of a cylinder of length $L$.
The minimal radius  $R_{\rm m}$ of this
cylinder is not determined by the geometry of the system, rather it
should be generated dynamically. If we assume that  $R_{\rm m}$ is
large  enough to apply CFT formulas, we get an asymptotic expansion
similar to eq.(\ref{ft})
\eq
\langle W_xW_{x+L}\rangle\sim{\rm e}^{\left(-\sigma2\pi
R_{\rm m}L+{\tilde c}L/12R_{\rm m}
\right)}~~,
\label{cft}
\en
where the first term at the exponent is the usual area term and the
second one is the universal quantum correction. Comparing eq.(
\ref{glueball}) with eq. (\ref{cft})  yields
\eq
m_G\,\simeq\, 2\pi\sigma R_{\rm m}-\frac{D-2}{12R_{\rm m}}~~~.
\label{mara}
\en
In the bosonic string picture there is no natural lower bound for
$R_{\rm m}$: the minimal area is obtained for $ R_{\rm m}\to0$ ,
where the quantum contribution diverges. On the contrary,  the self-
avoiding string picture we are describing gives obviously
$R_{\rm m}\simeq R_f$, otherwise there is an overlapping of the
colour flux tube. From eqs. (\ref{rf}) and (\ref{mara}) we get
\eq
\frac{m_G}{\sqrt{\sigma}}\,\simeq
\sqrt{D-2}\frac{3\pi^2-2}{6\sqrt{\pi}}~~~.
\label{mass}
\en

In Table I the values of the observables we have determined in
 eqs. (\ref{tc}),(\ref{rf}) and (\ref{mass}) through our
 asymptotic effective string scheme, are compared with the
corresponding data from numerical simulations on lattice
gauge theories with various gauge groups.
The agreement to the numerical simulations on LGT is in general rather
good. The fact that the lowest observed glueball mass for 3D $SU(2)$ LGT
is much larger than the lower bound fixed by the string might indicate
that in this case the ground string state is decoupled.
It would be interesting to do new numerical simulations in order to
complete the table and to  test the universality of these string
formulas with other  gauge groups.


\begin{thebibliography}{9}

\bibitem{no}H.B. Nielsen and P. Olesen, \NP {B61} (1973) 45;
G.'t Hooft, \NP {B72} (1974) 461
\bibitem{lu} M. L\"uscher, Nucl. Phys. { B180}[FS2] (1981) 317
\bibitem{david}F. David, Europhys. Lett.{ 9} (1989) 575
\bibitem{sed}A.G. Sedrakyan, \PL{ B260} (1991) 45
\bibitem{noi2} M. Caselle, R. Fiore, F. Gliozzi and M. Pri\-ma\-vera,
Phys. Lett. { 200B} (1988) 525
\bibitem{noi3} M.Caselle, R.Fiore and F.Gliozzi,
Phys. Lett. { 224B} (1989) 153  and Nucl.Phys B (Proc. Suppl)
{ 17} (1990) 545;
M.Caselle, R.Fiore, F.Gliozzi,  P.Provero and S. Vinti,
Int.\ J.\ Mod.\ Phys. { 6} (1991) 4885
\bibitem{cb}M.Caselle, R.Fiore, F.Gliozzi, P.Provero and S. Vinti,
 \PL {B272} (1991) 372
\bibitem{sve} B.Svetisky and L.G. Yaffe, \NP{ B210 [FS6]} (1982) 423;
 C. Borgs, \NP{ B261} (1985) 455
\bibitem{tc}M.Caselle and F.Gliozzi, \PL{ B273} (1991) 420
\bibitem{app}F. Gliozzi, lecture given at the XXXII Cracow School of
Theoretical Physics, June 1992, to be published on Acta Phys. Pol.
\bibitem{saw}M.Caselle and F.Gliozzi, \PL { B277} (1992) 481
\bibitem{olesen}P. Olesen, \PL{ B160} (1985) 144
\bibitem{df}P.Di Francesco, H.Saleur and J.B.Zuber, \NP { B285} (1987) 454
\bibitem{fp}M.Flensburg and C.Peterson, \NP {B283} (1987) 141
\bibitem{dg}A.Di Giacomo, M.Maggiore and S.Olejnik,
\PL {B236} (1990) 199
\bibitem{fh} J.Fingberg, U.Heller and F.Karsch, preprint BI-TP 92-26,
HLRZ-92-39, FSU-SCRI-92-103 (1992)
\bibitem{vb}P.van Baal and A.S. Kronfeld, Nucl. Phys. { B}
(Proc. Suppl.) {9} (1990) 227
\bibitem{te} M. Teper, Contribution to this conference
\end{thebibliography}
\end{document}